# A Construction Kit for Efficient Low Power Neural Network Accelerator Designs

Petar Jokic, Erfan Azarkhish, Andrea Bonetti, Marc Pons, Stephane Emery, and Luca Benini

*Abstract*—**Implementing embedded neural network processing at the edge requires efficient hardware acceleration that couples high computational performance with low power consumption. Driven by the rapid evolution of network architectures and their algorithmic features, accelerator designs are constantly updated and improved. To evaluate and compare hardware design choices, designers can refer to a myriad of accelerator implementations in the literature. Surveys provide an overview of these works but are often limited to system-level and benchmark-specific performance metrics, making it difficult to quantitatively compare the individual effect of each utilized optimization technique. This complicates the evaluation of optimizations for new accelerator designs, slowing-down the research progress.**

**This work provides a survey of neural network accelerator optimization approaches that have been used in recent works and reports their individual effects on edge processing performance. It presents the list of optimizations and their quantitative effects as a construction kit, allowing to assess the design choices for each building block separately. Reported optimizations range from up to 10'000x memory savings to 33x energy reductions, providing chip designers an overview of design choices for implementing efficient low power neural network accelerators.**

## I. Introduction

Machine learning (ML) algorithms, especially neural networks (NN), have been widely used to provide smart systems with complex data analysis capabilities like visual object detection [1] and audio key-word spotting [2]. While NNs enable algorithm developers to implement difficult-to-model tasks, given that sufficient training data is available, their computational complexity is challenging the design of processing hardware. Miniaturized and battery-powered ML applications thus require high computational throughput while being limited to low power consumption, rendering computing efficiency a key design objective for low power ML accelerators. The rapid progress of ML algorithm research further challenges designers to quickly adopt newly introduced network features, requiring fast development times. While fully-connected (FC) and convolutional neural networks (CNN) like AlexNet [3] have dominated the field during the past decade, residual networks (ResNet) [4], recurrent NNs (RNN) and derivations like dense CNNs [5] have gained importance, claiming ever-improving algorithm performance. What remains constant are the working horses of NNs, namely parallelized multiply-and-accumulate (MAC) operations.

Improving computing efficiency has been a key driver for the invention of laptops (~2000), smartphones (~2010) and high-performance server operation [6]. The current decade (~2020) strives for ubiquitous smart devices like wearables and internet-of-things (IoT) nodes [7], capable of processing sensory data. Performing data analysis on the sensor nodes at the edge of a connected network, so-called edge processing [8] or edge intelligence [9], can significantly reduce latency and power consumption but requires efficient ML accelerators.

Thus, edge processing is increasingly used in applications where long battery lifetimes are mandatory: e.g. in smart glasses with object detection [10], face detection [11], or hand-gesture and speech recognition [12], as well as smart cameras with automatic acquisition using scene classification [13], smart doorbells with face recognition [14], or tools that help blind people read texts and recognize people [15]. Many more could benefit from edge ML in the future [16, 17]. An overview of ML applications that are feasible on current hardware platforms is summarized in [18], illustrating the challenge of the limited edge processing power budget (<1W).

Existing ML accelerator chips cover the application domain from ultra-low power (ULP) and low-complexity processing, implementing 18uW key-word spotting with 105kB on-chip memory [19], to high-throughput server-grade acceleration, provided by chips like Google TPUv3 [20] or Graphcore IPU [21], consuming more than 100W. To identify relevant edge ML accelerator designs among the vast number and diversity of publications proposing efficient implementations, quantitative surveys are necessary. However, existing surveys often only provide qualitative comparisons or benchmark implementations on a system-level, obscuring the individual effects of each employed optimization technique. Comparing these optimizations is essential for motivating design choices during the development of new accelerators, currently requiring time-consuming literature research.

This work summarizes and, for the first time, quantitatively compares design optimizations of existing NN accelerators for

This work was supported in part by the EU project ANDANTE under Grant 876925 and in part by the Swiss National Science Foundation (SNSF) BRIDGE under Grant 40B2-0_181010.

Petar Jokic is with the Swiss Federal Institute of Technology, ETH Zurich, 8092 Zurich, Switzerland and CSEM SA, 8005 Zurich, Switzerland (email: petar.jokic@csem.ch). Erfan Azarkhish, Andrea Bonetti, Marc Pons, and Stephane Emery are with CSEM SA, 8005 Zurich, Switzerland (email: <first name>.<last name>@csem.ch). Luca Benini is with the Swiss Federal Institute of Technology, ETH Zurich, 8092 Zurich, Switzerland and the University of Bologna, 40126 Bologna, Italy (email: lbenini@iis.ee.ethz.ch).

tiny (<10mW) and edge (<1W) processing applications. It is presented as a construction kit, listing optimization options for each building block along with their reported quantitative effects, enabling application-specific integrated circuits (ASIC) designers to evaluate and assess optimizations for new implementations. Fig. 1 illustrates the covered building blocks on a generic end-to-end edge processing system and localizes optimizations within the edge ML design flow.

The paper is organized as follows: Section II presents related surveys that complement this work. Section III introduces basic notations used throughput the paper, followed by an overview of architectures in Section IV, technological optimizations in Section V, dataflow optimizations in Section VI, data handling optimizations in Section VII, computation optimizations in Section VIII, and finally the quantitative comparison of all relevant optimizations in Section IX.

## II. Related Work

Various surveys have been conducted to summarize existing NN accelerator designs and implementations, explaining their techniques, and comparing their system-level performance.

To keep track of the vast number of academic and industrial ML hardware accelerators proposed every year, a periodically updated online list [22] is provided, listing the main performance metrics for each chip to compare them in terms of power, throughput, and computational efficiency. A similar survey in paper form was presented in 2019 [23] and updated in 2020 [24]. It lists the computational performance and precision of academic research works and commercially available devices. A survey from 2017 claims to cover the past 35 years of works in neuromorphic computing, listing more than 2600 references that accelerated the research field since the 1980s [25]. It covers models, learning approaches and hardware ranging from analog to digital implementations, covering programmable FPGAs and custom ASICs.

Sze et al. [26] provide a thorough survey on efficient processing of deep NNs (DNN), covering historic aspects, common layer types, training frameworks and popular datasets, extending their previous work [27]. The sections on hardware platforms and energy efficient dataflows, are motivated in their preceding work [28], proposing edge processing for extracting meaningful information to reduce the extreme amount of data produced by the ever-increasing number of sensors in connected devices. A similar summary is presented in [29] and more FPGA-focused in [30].

Another well-structured and exhaustive survey on DNN acceleration is presented in [31], covering existing hardware acceleration approaches including software optimizations, and a chapter on security of DNN approaches and their benchmarking. Survey [32] presents a broad overview of the ML field, focusing on big data, training techniques, and applications. A similarly broad view, additionally covering the transition from modelling biological NNs to implementing artificial NNs in hardware is presented in [33], focusing on novel memories and their use-cases in the field.

In the survey of [34], various ML accelerators and processing blocks are presented and compared to their research group's own works. They list neuromorphic processors, including spiking NN engines, ranging from fully digital to fully analog computations, and discuss possible future directions. Survey [35] also discusses the architectures of selected DNN accelerators and explains their working principles and supported network architectures.

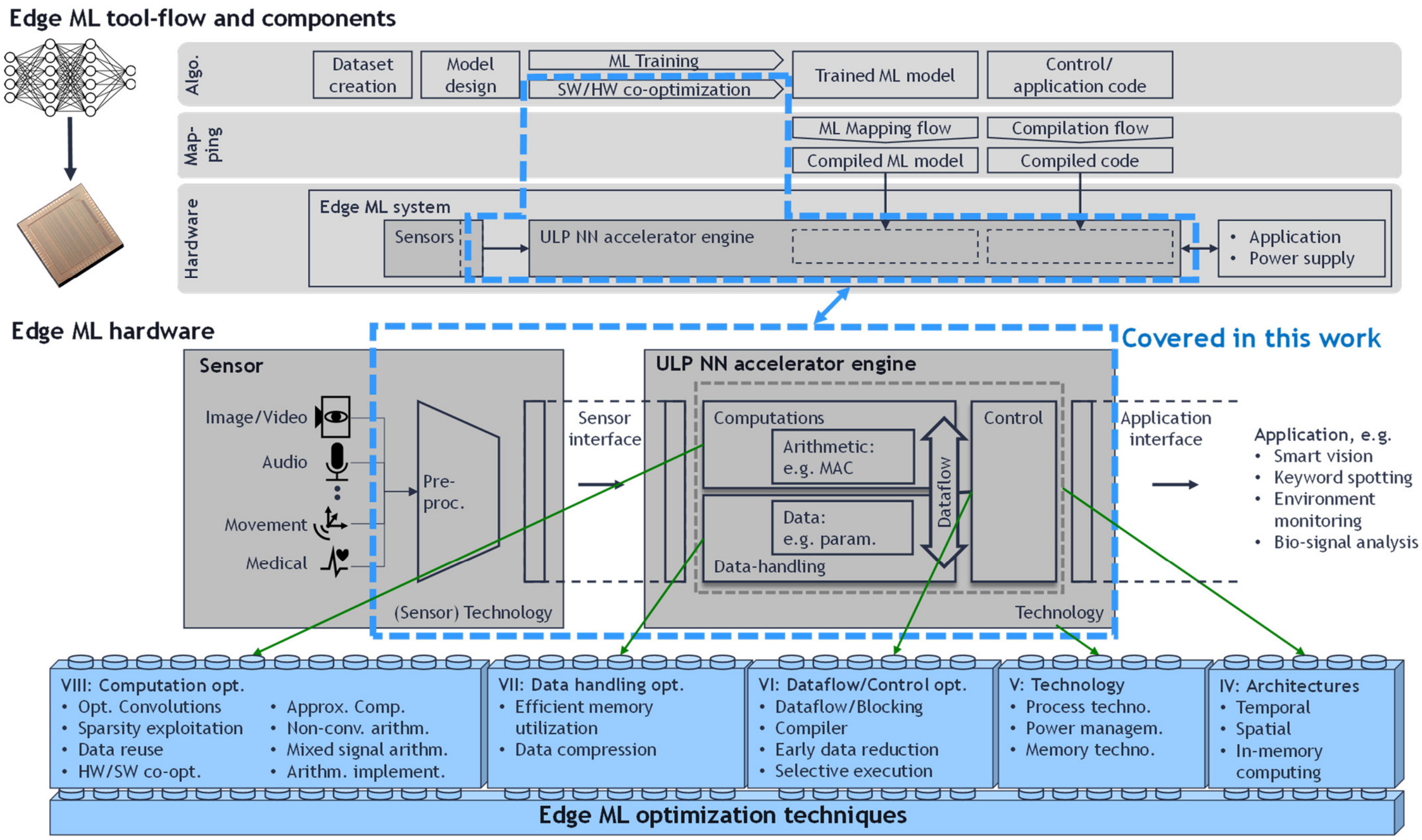

Fig. 1 Overview of the edge ML tool flow with a summary of the discussed hardware optimizations for NN accelerators (with chapter numbers of the paper).

Surveys [9] and [36] present an extended view on edge intelligence, covering pure edge processing and combinations with cloud processing. They discuss related optimization strategies, namely compression and early model exit.

While the listed related works give an excellent overview of existing ML accelerators and optimization techniques, none of them attempted to quantitatively compare used optimization approaches, as covered in this work, allowing designers to evaluate and assess optimizations for new implementations.

This work focuses on (deep) NN inference, noting that the field of ML is much broader, containing other approaches like support vector machines, decision trees and many others.

## III. Construction Kit Foreword

ML accelerators often combine multiple optimization techniques, as shown in TABLE I, summarizing a selection of relevant accelerator chips from the past six years. This complicates the assessment of individual optimizations as their effects are obscured by system-level benchmarking. Thus, we only add individual optimizations to the comparison and only if sufficient quantitative information is reported. We focus on five performance indicators, namely 1) energy/ power, 2) area (cost), 3) memory size, 4) computational throughput, and 5) impact on algorithmic accuracy.

In the following sections, we briefly discuss the importance of a system-level view for meaningful optimization evaluations and introduce some basic performance indicators and notations that are used throughout the paper.

TABLE I
Selected ML accelerator chips from last years

| Work (Year) | Optimizations | Throughput [GOPS] | Efficiency [TOPS/W] |
|---|---|---|---|
| ShiDianNao (2015) [159] | Local data reuse | 194 (16b) | 0.606 |
| EIE (2016) [139] | Sparsity, weight sharing, compression, zero skipping | 102 (16b) (~3'000 *) | 0.17 (5.0 *) |
| Envision (2017) [90] | Multi-precision, DVFAS, body biasing | 76 (4b) | 10 |
| Eyeriss (2017) [136] | Local data reuse, zero compression, zero skipping | 84 (16b) | 0.166 |
| YodaNN (2018) [175] | Binary weights, standard-cell memory, voltage scaling | 1'500 (1b \| 12b w. \| a.) | 1,1 |
| UNPU (2018) [202] | Multi-precision, LUT-based bit-serial MAC (1-16b) | 346/7372 (16b/ 1b \| 16b w.\|a.) | 3.1/50.6 |
| Eyeriss v2 (2019) [137] | Local data reuse, sparsity, compressed computing | 202 (8b) | 0.963 |

* including skipped operations

### A. System-level view

Edge ML devices process sensory data onboard, communicating with sensors (and memories) for subsequent analysis in a ML engine. Regardless of this fact, publications on low-power ML accelerator designs often neglect the impact of such off-chip communication on system-level power consumption and performance. Analyzing system-level power-breakdown helps identifying power-dominating subsystems, allowing to optimize them based on the Pareto principle. Fig. 2 shows the power breakdown of four mobile system implementations: two smartphone analyses were taken from a smartphone battery usage review [37], showing dominating communication power, while two IoT nodes were evaluated in a visual presence detection system [38] and an always-on face recognition system [39], reporting dominating sensor and processing power. While core optimizations would only enable marginal system improvements in the first three systems, the fourth example shows a typical edge ML IoT node with processing-dominated power distribution, enabling significant system improvements through core optimizations.

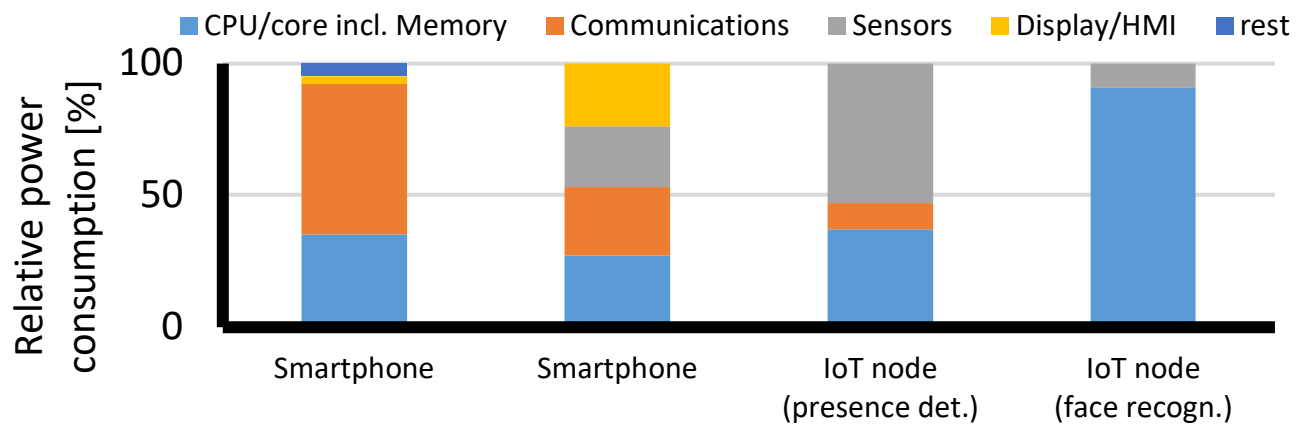

Fig. 2 Power break-down of mobile/edge systems (smartphones [37], presence detection IoT node [38], and face recognition IoT node [39]).

### B. Meaningful performance indicators and metrics

To identify useful optimization strategies, relevant performance indicators and benchmarks must be chosen. Parameter quantization, for example, can easily reduce memory size but also heavily impacts accuracy [40], [41]. Similarly, the throughput is subject to large variations across different workloads as pointed out in [23], reporting 20x lower throughput than stated in the datasheet. Thus, application-relevant benchmarks are indispensable. For a fair cross-device comparison, standard ML benchmarks have been created for smartphones [42], for general purpose devices (MLPerf) [43], and are now adopted to edge devices (TinyMLPerf) [44]. This is similar to microcontroller benchmarks (e.g. CoreMark [45]).

Roofline models [46] are used to visualize the architectural boundaries of a system's operating points, namely the memory bandwidth and the peak computational throughput, as illustrated in Fig. 3. Depending on the operating point, the system operates in a memory- or a computation-bound region.

NN accelerator performance metrics are often limited to the peak throughput, in tera operations per second (TOPS), and the computational power efficiency (TOPS/W). Fig. 4 illustrates that these metrics can be misleading in edge applications, operating in the low power region where extrapolating the efficiency becomes inaccurate (idle power).

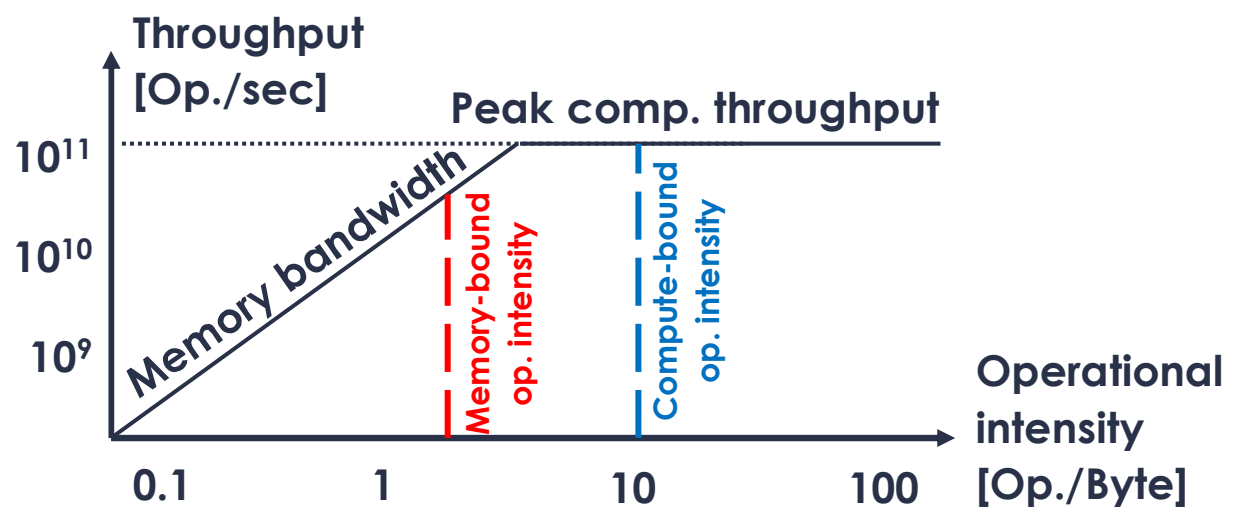

Fig. 3 Roofline plot showing the operation point, constrained by the available memory bandwidth and the computational throughput.

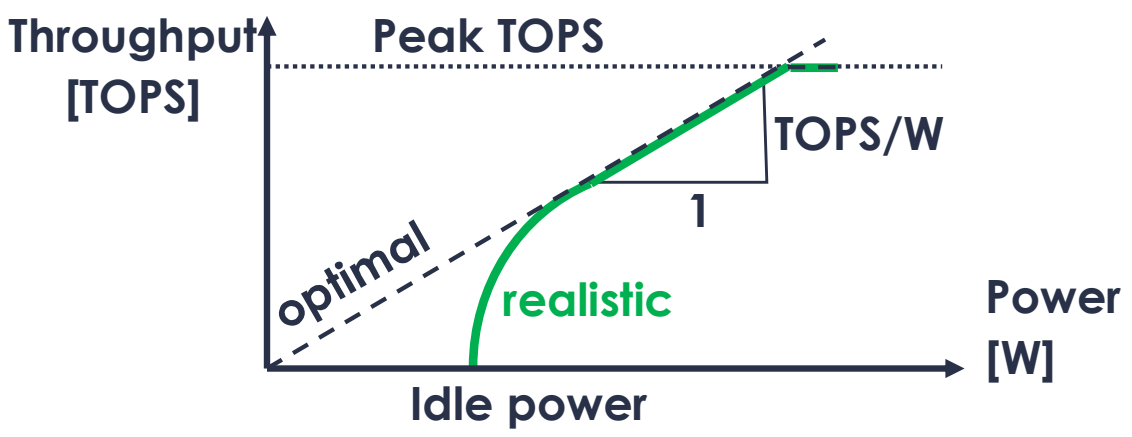

Fig. 4 NN processing throughput versus power consumption.

### C. Neural network notations

Fig. 5 illustrates a generic NN layer along with the notations of layer dimensions that we use throughout the paper. The input activations $in$ have $C_{in}$ input channels and extend $X_{in} \cdot Y_{in}$ in spatial dimensions. All $C_{out}$ kernels contain $C_{in} \cdot k_x \cdot k_y$ parameters and generate output activations $out$ of dimension $X_{out} \cdot Y_{out}$ with $C_{out}$ output channels. While a generic convolution layer is shown here, it also covers dense (FC) layers, which have restricted dimensions $k_x = k_y = X_{in} = Y_{in}$ and $X_{out} = Y_{out} = 1$. Other layer types can be described using the same notation: e.g. ResNets have additional bypass inputs $in_{by}$ of the same dimension as $in$ and are added point-wise. Depth-wise separable CNNs split kernels in the $C_{in}$ dimension, yielding $C_{out} = C_{in}$, avoiding cross-channel links.

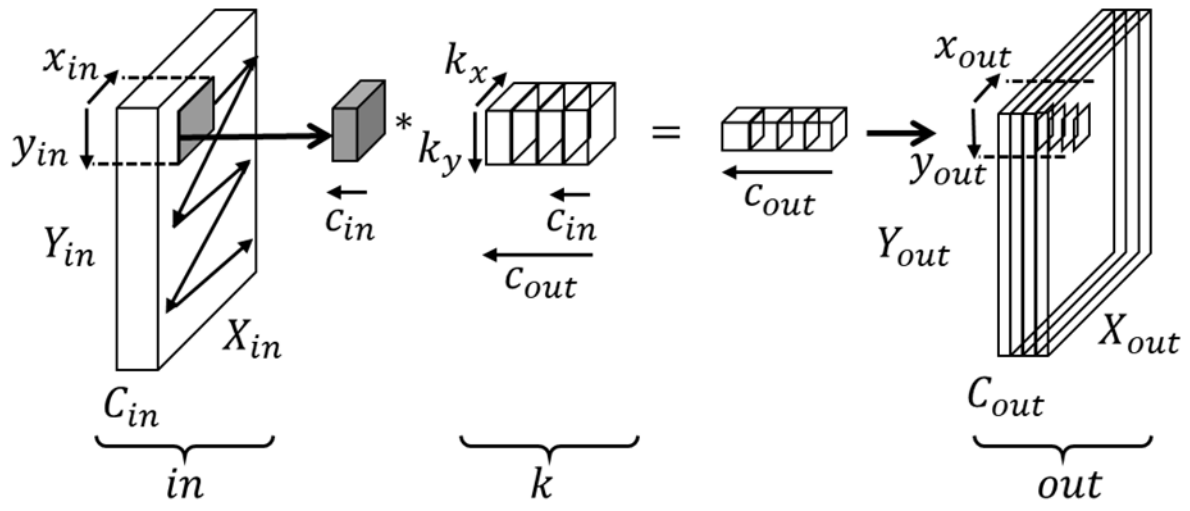

Fig. 5 Generic CNN layer with dimension notations.

## IV. Hardware architectures

Two main architectural concepts are used in today's NN accelerators [27]: temporal and spatial architectures. This chapter discusses them and adds emerging in-memory computing as a third architecture, offering a distinct data flow.

### A. Temporal architecture

Temporal architectures comprise, among others, the wide-spread central processing units (CPU) and graphic processing units (GPU), featuring a central control unit that schedules tasks and distributes computations across arithmetic logic units (ALU). Data is moved from the memory to the ALU and back, offering a high parallelization potential using single instruction, multi-data (SIMD) instructions for vector processing. This is based on the traditional Von Neuman architecture [47], characterized through a generic single instruction, single data (SISD) processing scheme that stores both instructions and data in the memory. It typically features a control unit that reads instructions and data from the memory, an ALU to perform the operations, and a data bus to communicate across blocks and through a peripheral interface.

Multi-issue processors [48], e.g. super-scalar processors, extend this concept by enabling multiple parallel operations through a single instruction, reducing the control overhead, to increase the operational intensity (see Fig. 3).

### B. Spatial architecture

In contrast to temporal architectures, spatial architectures allow their arithmetic units, often called processing elements (PE), to move data between neighboring PEs, allowing to reduce memory accesses by employing local buffers.

Systolic arrays are pipelined 2D spatial architectures that enable data reuse across neighboring PEs. This can be exploited for implementing efficient general matrix multiplication (GEMM) through contraction ("systole" in old Greek) of computations, reusing results from adjacent nodes, and thus minimizing memory accesses. This is exploited in many prominent accelerators like Google TPU [49] and Eyeriss [50]. Replacing single-PE systolic arrays with tensor-PEs, each one computing an entire matrix multiplication per cycle, further allows reducing area and power in a 16nm process by 2.1x and 1.4x, respectively [51]. Increased intra-PE data reuse and fewer pipeline buffer registers enable this improvement but require efficient load distribution to avoid low PE utilization. SCALE-Sim [52] provides a simulation tool to evaluate such design parameters, comparing different dataflows and arrays.

### C. In-memory computing

The motivation behind in-memory computing, or compute-in-memory (CIM), is the data-intensive nature of neural network inference, requiring high memory bandwidths which result in memory-dominated performance [53], the so-called memory wall [54]. To mitigate this, computations can be directly performed in the memory, where high data access rates are available at a much lower power cost. While the computational efficiency can be largely increased with this approach, it increases the overhead in the memory and limits the flexibility. Combined with analog memory types, the efficiency can be further improved by computing directly in the analog domain. Analog CIM computes matrix multiplications by breaking them down into vector dot products [55], multiplying analog input voltages (activations) with NVM cell conductances (weights) and accumulating the resulting column current (MAC result). Special design considerations to achieve high accuracy inference along with high computational efficiency are discussed in [53].

SRAM-based CIM is presented in [56], using a 55nm 8T 3.8kb SRAM macro with support for 1-4b input activation and 1-5b weights. The 130nm circuit in [57] demonstrates simple down-sampled MNIST computations, reporting 13x system energy reduction for 1b weight and 5b activation precision compared to a traditional (out-of-memory) computation. CIM integrated in the analog SRAM periphery is evaluated in [58] on a simulated 65nm process, reporting 4.9x system energy efficiency and 2.4x throughput improvement compared to digital processing. A 384kb SRAM-based 8b precision CIM in 28nm [59] is demonstrated to have 28% area overhead compared to a pure SRAM array while achieving up to 22.75TOPS/W throughput.

Non-volatile memory (NVM)-based CIM implementations are summarized in [55]. Their previous survey [60] provides a comprehensive list of the utilized emerging NVMs, illustrating that NVM can increase bit density and reduce leakage compared to SRAM and possibly store multiple bits per cell. In [61] an MRAM-based 54 x 108 CIM crossbar is presented in a 180nm process (MRAM on top of CMOS circuit). Over-lifetime variations of up to 4.2% and device-to-device variations of 4.5% have been reported, requiring special considerations (e.g. [62]). Among many other RRAM works, a 1Mb multibit CIM on a 55nm process [63] and a 128 x 64b CIM array [64] on a 90nm process are presented. NVM-based CIM is an active field of research with various directions, for example 3D CIM architectures [65], reporting up to 28.6x higher energy efficiency compared to 2D chips. A CIM benchmarking tool [66] further reports advantages of NVM- over SRAM-based CIM implementations in a 32nm process, while 7nm SRAM CIM still outperforms any NVM-based work in throughput and both area and energy efficiency.

Note that also DRAM has been used for CIM [67], however, targeting high performance server applications which goes beyond the scope of this work.

# V. Technology

Integrated circuits (IC) for NN accelerators are strongly influenced by the underlying semiconductor technology. This chapter introduces common process technologies and related power optimization techniques, followed by an overview of memories, which are often linked to process technologies.

## *A. Semiconductor process technology*

Annual improvements in semiconductor manufacturing technology have been a major driving force in the chip industry as indicated by the (now saturating) Moore's law [68, 69], empirically predicting the doubling of component density on chips every 1-2 years. However, the smaller process nodes increased the static power consumption, resulting in the end of Dennard scaling [70]. This related heuristic scaling trend factor describes the annual shrinking of the minimum feature size in silicon chips and related power savings, culminating in the so-called power wall [71], limiting process improvements because the increased power density has approached physical limits of silicon-based circuits over the past few years.

This process scaling enables significant power reductions [72], as it yields lower supply voltages and smaller switching capacitances. However, improved process technologies are often linked to a significant cost increase and might not support all features of older technologies (e.g. special memory types, photo diodes, etc.). Multi-die solutions can mitigate this problem, exploiting the properties of multiple processes across the dies, each one optimized for a specific target like reduced cost, specialized memory support, or high logic density [73, 17]. Combining multi-process solutions with 3D die stacking additionally provides short communication paths and increased densities, as shown in the 8-die-stacked NN accelerator with 96MB of memory [74].

The following sub-sections give a brief introduction of the main semiconductor process technologies used today as they will be referred to throughout the paper. We limit the scope to complementary metal-oxide-semiconductor (CMOS) technology, which dominates digital designs and refer the interested reader to the annual IEEE white paper on future directions of semiconductor technologies (e.g. the 2020 update [75]) for a look into possible future directions.

### *1) Bulk*

Bulk technology is based on standard silicon wafers and mainly evolves through spatial scaling. However, these annual scaling improvements are slowing down due to increasing difficulties with electrostatic and short-channel effects [76].

Deeply depleted channel (DDC) technology improves bulk technology by introducing multiple vertical doping regions in the channel, forming a threshold setting region and a bias-controllable screening region [77]. This enables reduced supply voltages, low leakage, low process variation as well as improved body biasing characteristics, allowing to dynamically adjust the threshold voltage of the transistor.

### *2) FinFET*

The introduction of 3-dimensional gates in so-called FinFETs improved on the performance of bulk CMOS by enabling lower supply voltages, reducing power consumption, as shown for 22nm tri-gate FinFET [76] and later 14nm FinFET [78]. However, the improved performance comes at a higher production cost due to the complex 3D structures, requiring more masks than the bulk technology [79, 80, 81].

### *3) FD-SOI*

Fully depleted silicon-on-insulator (FD-SOI) technology employs very thin insulating layers in the substrate of the transistors, reducing leakage. In [80] a 22nm FD-SOI technology is presented, achieving on par power and performance efficiency compared to 16/14nm FinFET technology while being lower cost due to the use of a planar processes, requiring fewer masks. Thus, FD-SOI is more suitable for low-end mobile and IoT applications where cost is an important factor. A detailed comparison between FD-SOI and FinFET is provided in [79], reporting superior performance of FinFETs in terms of power, delay, and density, for which FD-SOI can compensate through body-biasing (BB). BB allows to dynamically adjust the threshold voltage, boosting speed with a forward body bias or reducing leakage using reverse body biasing (higher threshold).

### *4) Specialized processes*

Recent advances in materials and manufacturing technologies have enabled the integration of novel memory technologies (both volatile and non-volatile) close to processing logic. They offer advantageous characteristics that go beyond transistor density scaling. However, most of them require special process technologies, making them more difficult to integrate within the widely available processes. More details on novel memories can be found in Section V.C.

## *B. Power management*

The power consumption of ICs [71] can be decomposed into dynamic and static power. Dynamic power is described in Equation (1), taking the circuit switching operations into account. Variable $U$ is the supply voltage, $C$ the switching capacitance, $\alpha$ the switching activity and $f$ the frequency of the circuit. Static power, often also called leakage power, is described in Equation (2), reflecting the current consumption $I_{leak}$ when no switching takes place.

$$P_{dynamic} \sim \frac{1}{2} \cdot U^2 \cdot C \cdot \alpha \cdot f \quad (1)$$

$$P_{static} \sim U \cdot I_{leak} \quad (2)$$

Based on these equations, various optimizations have been proposed to reduce the overall power consumption, using the supply voltage and the frequency as control knobs. The most prominent techniques are listed in the following section, namely sub- and near-threshold operation, adaptive body biasing (ABB), and dynamic voltage and frequency scaling (DVFS). If the application permits, duty cycling (pausing the operations to reduce the switching activity $\alpha$) and power gating (to reduce static current $I_{leak}$) can be used to further reduce the power consumption.

*1) Sub- and near-threshold operation*

Sub- and near-threshold operation exploits the supply voltage knob, operating transistors below or close to their threshold voltage, respectively, to reduce power consumption [82]. This enables to reduce dynamic power quadratically and static power linearly, as shown in Equations (1) and (2).

However, these savings come at the cost of slower transistor operation, limiting the application frequency. In embedded IoT applications, energy is usually more important than power consumption, as it directly determines the lifetime of the battery. Thus, the minimum energy point (MEP) is identified for each application by adjusting the supply voltage such that the total energy for a specific workload is minimal. Sub- and near-threshold operation extends the supported supply voltage range, reaching the MEP for a large set of workload scenarios.

Lowering the voltage implies higher sensitivity to process variations, which must be carefully evaluated during the design phase to ensure robustness under all operating conditions. Special layout considerations and compensation techniques (see ABB and DVFS) can reduce the effects.

The 180nm sub-threshold standard cell library developed in [82] demonstrates an extended 0.4-1.0V supply voltage, reducing power by 5x compared to a standard low power library at 1.0V. Their follow-up work presents 1kb sub-threshold SRAM [83] and a 32b microcontroller [84] design, achieving 0.84-3.2nW (3.8x) power scalability for 0.27-0.6V voltage scaling and 7x power scalability for 0.37-1.8V voltage scaling, respectively.

*2) ABB*

Adaptive body biasing (ABB) [85] adjusts the bias voltage of the transistor body to control its threshold voltage, influencing its speed and power consumption as discussed in Section V.B.1). FD-SOI and DDC can fully exploit BB, while the effect in bulk technology largely depends on the design parameters. Adapting the BB to the operating point enables to reduce the adverse effect of sub-threshold operation on timing across process and temperature variations (e.g. worst case corner distance from 21x to 0.2x [86]). This allows to implement a wide range of timing-clean operating points from fast to slow and low power. Note that increasing the speed through a strong forward body bias also increases the leakage.

Publications on ABB report 30x frequency and 20x leakage scaling on a RISC microcontroller core and SRAM [85].

*3) DVFS*

Dynamic voltage, frequency, (and accuracy) scaling (DVF(A)S) allows to trade-off speed against power consumption through supply voltage scaling. The basic principle was already evaluated in the 1990s, allowing CPUs to lower the frequency and voltage for low intensity tasks, reporting power reductions of 1.05-4x [87], and 9.2x [88].

DVAS [89] extends the principle to dynamic accuracy scaling through adjustable arithmetic bit-widths, allowing for lower supply voltages due to shortened critical paths. Demonstrated on a simulated 40nm 16bit array multiplier, DVAS achieves 11.7x energy reduction for scaling down to 8b precision, noting that the critical path length is reduced by 40%. Pipelined architectures require bypassable pipeline registers to evenly distribute the path length reductions, adding area and energy overhead. On a 16b multiplier, 11.1x energy reduction is reported for 8b operation at 8% energy overhead.

Envision [90] combines accuracy scaling with DVFS on a 28nm process, running at constant throughput while scaling precision from $1 \cdot 16$b operations to $4 \cdot 4$b operations at a quarter of the 16b frequency. Combined with body biasing to reduce leakage at low frequencies, power is reduced by 25x.

## C. Memory

Data handling is dominating today's accelerator power consumption and area, requiring careful selection of the memory type and the access strategies. Memory access energy is subject to large variations across memory types and sizes, as summarized for a standard 45nm SRAM [72, 91] in TABLE II. DRAM is reported to have 200x higher access energy than a small 8kB SRAM (for 64b word width) [72].

Fig. 6 shows an overview of the existing memory types, distinguishing storage (e.g. non-volatile hard disk) and memory (e.g. RAM) due to their significant difference in access times, density and power consumption [92]. Systems with long idle/sleep phases might not be dominated by access power but (idle) leakage power. In that case, rare accesses to (power-intensive) non-volatile memories could be cheaper than retaining data over a long period of time, constantly consuming leakage power.

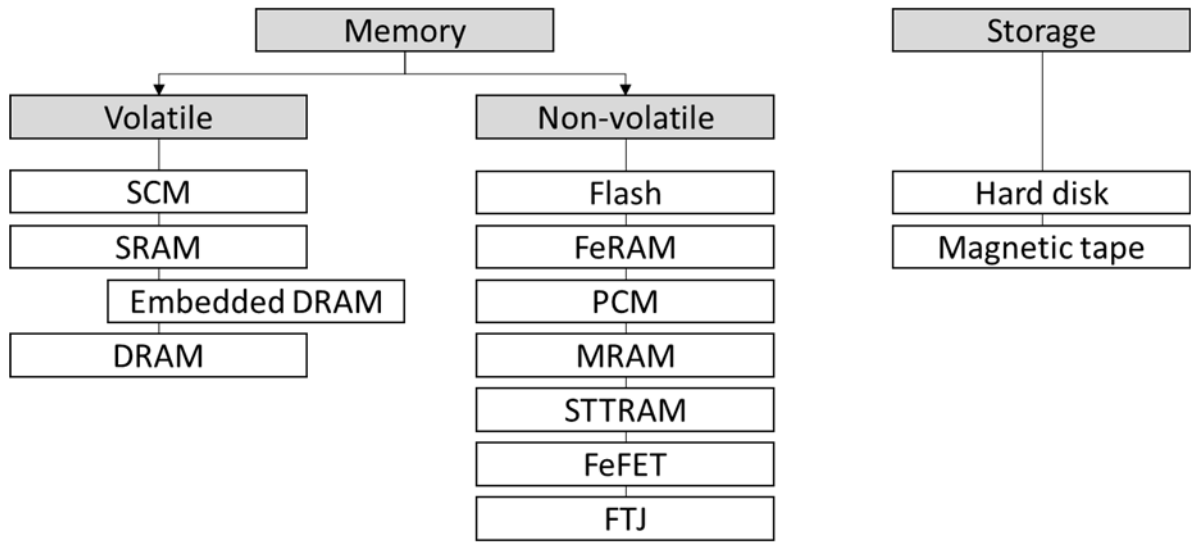

Fig. 6 Overview of memory technologies.

TABLE II
ENERGY PER MEMORY ACCESS IN 45NM SRAM

| Size(kB) | Access energy per 16b word [pJ] | | | |
|---|---|---|---|---|
| | 64b | 128b | 256b | 512b |
| 1 | 1.2 | 0.93 | 0.69 | 0.57 |
| 2 | 1.54 | 1.37 | 0.91 | 0.68 |
| 4 | 2.11 | 1.68 | 1.34 | 0.9 |
| 8 | 3.19 | 2.71 | 2.21 | 1.33 |
| 16 | 4.36 | 3.57 | 2.66 | 2.19 |
| 32 | 5.82 | 4.8 | 3.52 | 2.64 |
| 64 | 8.1 | 7.51 | 5.79 | 4.67 |
| 128 | 11.66 | 11.5 | 8.46 | 6.15 |
| 256 | 15.6 | 15.51 | 13.09 | 8.99 |
| 512 | 23.37 | 23.24 | 17.93 | 15.76 |
| 1024 | 36.32 | 32.81 | 28.88 | 25.22 |

The range of available memory technologies is rapidly increasing, being an active area of research. A recent overview of novel memories used in neuromorphic computing [93] notes that the memory technologies mostly differ in their writing speed, while the reading process is dominated by the sensing circuit interfacing the memory cells. Survey [92] provides an overview of recent non-volatile memories, focusing on PCM, STTRAM, RRAM, and FeFET. We summarize the working principles and extend it with the dominantly used SRAM and standard cell memories as well as quantitative optimization results from the literature.

*1) PCM*

Phase change memory (PCM) [92] is based on the heat-induced reversible phase transition of chalcogenides. It can switch between the low-resistance crystalline phase and the high-resistance amorphous phase, implementing a bipolar switch. The relatively large switching current requires powerful access circuits, dominating the size. Speed is limited by the transition from amorphous to crystalline phase while the reverse process dominates the power consumption. Endurance and speed are estimated around 1G cycles and <100ns, respectively [92]. TSMC presents a 1Mb PCM memory array in 40nm [94], reporting 300uA write current at 100ns write speed, achieving >200k cycles endurance.

*2) STTRAM*

Spin-transfer-torque RAM (STTRAM) [92] is based on a magnetic tunnel junction (MTJ) with two ferromagnetic layers separated by an ultra-thin tunnel oxide layer. It uses the spin transfer torque of spin-polarized electrons to change the resistance of the memory element, enabling non-volatile states. The access circuit dominates its size due to the relatively high writing currents but is still smaller than SRAM.

STTRAM is demonstrated on 7-8Mb arrays (industrialized 1Gb cluster [95]) in 22-28nm [96, 97], reporting densities up to 10.6Mb/mm$^2$ and write endurance of >1M-10G cycles.

*3) RRAM*

Resistive RAM (RRAM) [92] stores information by modulating the resistance of a metal oxide and is therefore often called memristor. A similar approach is conductive-bridge RAM (CBRAM), that forms a conductive metallic bridge in the on-state and interrupts it for the off-state. Endurance of 1M-1G cycles have been reported. However, there are tradeoffs between speed, power, and endurance.

Various RRAM implementations in 14-40nm have been shown [98, 99, 100, 101, 63], reporting 0.9-244.8Mb/mm$^2$ density at supply voltages down to 0.7V for 1Mb-32Gb sizes.

*4) FeFET*

Ferroelectric FET (FeFET) [92] employs a ferroelectric gate dielectricum that allows to change the resistance of the FET in a non-volatile fashion. This principle is similar to the Flash technology, that uses a floating gate instead. While the power consumption is low due to the low leakage through the gate oxide, the switching speed is high (~20ns). However, its endurance is relatively low, at 10k-100k cycles. Its similarity with standard CMOS transistors makes FeFET compatible with many standard processes.

In [102], a 10Mb FeFET memory array is implemented in a 22nm process, reporting 200k cycles endurance at 2.5-4.5V supply voltage. Their previous work [103] presents a 64 x 64b array in the same 22nm process and compare it to a 6T SRAM array of the same size, reporting 74x lower static power and >5.3x lower area for FeFET cells (without peripherals). Writing is 10x more energy-intensive and 10x slower while reading costs 1.6x more energy but is 1.5x faster.

A similar, but less mature, type of ferroelectric memory is ferroelectric tunnel junction (FTJ). The tunneling resistance of its ferroelectric layer between two metal electrodes can be adjusted through ferroelectric polarization reversal [92].

*5) SRAM*

Static random-access memory (SRAM) is the most often used on-chip memory as it can be easily implemented along with digital circuits. It features higher memory density than standard-cell memory as foundries use “layout pushed rules” to optimize SRAM bit-cell area beyond standard layout rules.

Standard SRAM bit-cells use 6 transistors (6T), but many larger cell structures have been proposed for power reductions. For low leakage operation a 7-transistor SRAM is presented in [17], reducing area by 18% and 50% compared to standard low leakage 8- and 10-transistor designs, respectively, while achieving similar performance and leakage power.

Power optimizations using non-uniform memory hierarchy in SRAM is presented in a 40nm NN accelerator [104], allowing up to 60% power savings when accessing the smallest instead of the largest level (32x smaller memory). The 67.5kB memory is split in 4 levels (1.5, 6, 12, and 48kB). SRAM access energy is shown to increase nearly linearly with the memory size above ~100kB [91] as shown in TABLE II.

Low leakage SRAM sleep-mode retention is presented on a 55nm process [105], reporting 26x lower retention power for a 16kB memory. Leakage is reduced by optimizing the design rules (increasing area by 2.7x) and the process corner. An additional sleep controller with a charge-pump for the retention voltage allows to power-gate the rest of the chip. Leakage is also reduced in [73], using 180nm low leakage 10T SRAM along with a 65nm 8T SRAM for dense scratchpad memory. The low-leakage memory consumes 4242x less standby power while being 11.3x larger in area.

To optimize the data access for 2D structures like in CNNs, a transpose SRAM has been proposed [106], allowing to selectively read a row vector or a column vector of data in parallel, reduce power consumption by 47%.

#### 6) SCM

Standard-cell memory (SCM) is implemented directly in digital logic using flip-flops or latches. This allows exploiting voltage scaling capabilities and avoids dependencies on vendor-specific memory generators.

A dedicated placement strategy for SCM (instead of standard logic place-and-route (P&R)) is presented in [107], reporting area and power savings on a 28nm process. Their experiments on 256b-32kb SCM macros show area reductions of >35% compared to standard P&R with access energy reduction of up to 65% for reading and 50% for writing. SCM macro sizes of up to 1kb are shown to be smaller than SRAM, but already 2-3x larger for 4kb macros (larger area of D-latch cell compared to 6T SRAM bit-cell starts dominating).

A BNN accelerator [108] with a hybrid memory consisting of 456kB SRAM and 8kB SCM demonstrates power savings of SCM compared to SRAM. It reduces the supply voltage to 0.4V when SCM is used, while SRAM requires 0.6V, leading 3x energy savings in 22nm post-layout measurements.

#### 7) DRAM

Dynamic RAM (DRAM) is a volatile high-density memory that stores information as a capacitor charge, which is periodically refreshed. DRAM is usually not compatible with standard logic processes, requiring separate DRAM dies. However, the hybrid memory cube (HMC) architecture [109] proposes high density DRAM access to standard logic processes through 3D stacking of DRAM dies on top of a logic die using through-silicon vias. HMC is implemented in TETRIS [110], providing DRAM access to a 45nm processing die. 3D DRAM is shown to consume 3.5-4.2x more energy compared to on-chip 256kB SRAM, but 1.5x less than a planar baseline DRAM at 4.1x higher throughput.

Embedded DRAM (eDRAM) [111] is a CMOS-compatible derivative of DRAM, targeting high density volatile memory. A 4-transistor 8kb eDRAM array is presented in 28nm [111], reporting 17% lower area and 23% lower static power consumption compared to 6T SRAM at equal voltage. The same author also evaluates a 2T design [112], showing slightly lower retention power than low power SRAM cells but 2.5-3.8x lower size for a simulated 28nm 4kb array.

#### 8) Flash

Flash memory [113] dominates NVM technology on the market, being embedded in most commercial chips where data retention during off-state is required. It can be implemented using standard logic process flows by adding a few additional masking layers, e.g. 3 extra masks on a 65nm process [114].

## VI. Dataflow and control optimizations

NNs have a relatively simple structure, their efficient implementation on hardware accelerators often complicates the dataflow. Efficient parallelization requires smart workload distribution among processing elements while ensuring coherent algorithmic functionality. Thus, this chapter discusses algorithm blocking optimizations, efficient scheduling, selective execution, and early data reduction.

### A. Dataflow and blocking

NN accelerators have optimized memory allocation and access patterns for efficient computation, splitting a task into smaller blocks that fit on the available resources. The utilized blocking (scheduling) strategy impacts the data access order and thus possible data reuse within the computation blocks. Higher reuse can reduce the number of memory accesses, decreasing the total power consumption [91].

#### 1) Layer-wise processing

Traditionally, NNs are processed layer-by-layer, also called layer-wise or layer-first approach. This enables the reuse of layer parameters (e.g. convolution weights), as they are repeatedly used across the layer. However, this also implies that at least one complete layer is always buffered in memory, as explained in more detail in section VII.A.

#### 2) Depth-first processing

The increasing size of feature maps (e.g. higher resolution images) requires large activation buffers to be allocated for layer-wise processing. However, networks can be processed in a "depth-first" streaming fashion instead [115], allowing each layer to buffer only a minimum set of input activations that are needed for computing the next set of output activations. Up to 200x memory bandwidth reduction or alternatively up to 10'000x memory space reduction is reported for this approach. In a follow-up work [116] depth-first processing is implemented on a FPGA and benchmarked on five models reporting throughput increase of 0.91-1.27x and memory bandwidth reduction of 3.9-81x.

A similar work on "fused layers" [117] observes that each output of a convolution layer only depends on a small region of input values. Tracking these dependencies back through multiple layers, results in a pyramid-shaped region. The paper proposes to compute the entire pyramid until the final output while only storing the computed intermediate features instead of buffering each complete. The additional cost for buffering intermediate results locally is traded-off against external memory accesses, achieving up to 95% reduction in off-chip memory traffic for running VGGNet-E.

#### 3) Loop ordering and optimization

Neural network can generally be described using nested loops, with the outer most one looping through the layers of the network. The ordering of these loops influences the possible parallelisms and the required memory size. To process larger networks on limited on-chip memory resources, loop tiling is used: the workload of each layer can be split into overlapping tiles, which are processed sequentially. Parallelization and data reuse can be increased by unrolling parts of the sequential loops and thus parallelizing their computations. Unrolling the entire kernel computation is shown in [118], achieving unprecedented power efficiency at the cost of larger area and limited layer size support.

Various tools have been proposed to optimize loop ordering [119, 91, 120, 121], improving energy efficiency and memory size.

### B. Compiler

Compilers and NN mapping tools translate the algorithmic representation of a trained network into machine code that is executed on the processing hardware as shown in Fig. 1. They can optimize dataflow strategies as listed in Section VI.A.

Commonly implemented microcontrollers provide specialized libraries to make NN processing more efficient. For example, ARM provides the CMSIS-NN library for its Cortex-M microcontrollers [122], supporting CNN, FC, and pooling layers, as well as 8b or 16b fixed point precision. Evaluated on a network running the CIFAR-10 task, a 4.6x improvement in throughput and 4.9x in energy efficiency is reported, compared to a DSP-functions-limited baseline code.

A biomedical signal analysis application [123] reports energy savings of 41.6% for enabling 8 core processing instead of single core, noting that the overhead of multi-core execution is fully compensated by efficiency improvements above a certain throughput. Block and memory power gating shows 16.8% energy savings but must be traded off against restart time and related energy and storage implication.

### C. *Early data reduction*

The high cost of data access during NN inference motivated to reduce the data flow from the sensor, condensing information early and thus minimizing costly data transfers.

One approach is to process the first layer(s) of a NN in the sensor itself. In [124] diffraction gratings above the pixels are used to optically detect Gabor filter-like patterns, as they are often found in the first layer of NNs. Evaluated on the MNIST and CIFAR-10 tasks, sensor communication bandwidth reductions of 10x are reported, with moderate accuracy impacts of -0.1% and -4.6%, respectively. However, the first layer of the employed LeNet-5 only accounts for 3.8% of all operations, rendering the computation reduction negligible.

Another study implements the first CNN layer in the optical domain using a controllable (grayscale) mask [125]. All filters (output channels) are displayed in the same plane, allowing the image sensor behind to capture all convolution results in parallel, forwarding them to the last layers implemented in the digital domain. Evaluated on MNIST and EMNIST tasks, operation reductions of 250x and 460x are reported while accuracy drops by 0.4% and 1.7%, respectively.

The analog nature of most sensor signals requires analog-to-digital conversion (ADC) which can be exploited by implementing matrix multiplication directly in the ADC [126]. An algorithmic reformulation is shown to implement a simple classification task using boosted linear classifiers, embedded in a single matrix transformation. The matrix multiplication is implemented in the feedback path of the SAR ADC, reporting 13x and 29x energy savings compared to SVM-based implementation with similar accuracy for ECG arrhythmia detection and 160x120 pixel gender classification tasks.

RedEye [127] implements an image sensor with analog on-die CNN processing capabilities on a simulated 180nm process. It uses SAR ADCs and tunable capacitors to implement weighted summation to mimic MAC operations, reporting 73% system energy reductions for running the first 1-5layers of an 8bit GoogleNet on the ImageNet task.

Early data reduction is also implemented in distributed computing [128, 129], splitting the DNN computation across the edge and the cloud to reduce costly data communications, latency and preserve privacy by keeping raw data at the edge.

Furthermore, application-specific early data reduction mechanisms exist: visual attention [10] is shown to reduce the object recognition workload in smart glasses by limiting the analyzed region to the detected eye-gaze direction.

### D. *Selective execution and early abortion*

This section presents techniques to dynamically adapt the network complexity to the input, enabling "simple" inputs to be analyzed with a fraction of the network capacity without decreasing the accuracy for complex ones. Average latency and power consumption can thus be reduced if simple inputs dominate the execution. Fig. 7 illustrates the techniques, covering a) hierarchically scalable effort [130], [90], b) early exiting [131], [132], and c) selective execution [133].

#### 1) *Hierarchically scalable effort*

Identifying early exits during training enables 2-6x latency reduction on MNIST and CIFAR-10 tasks [130]. A scalable-effort approach [134] proposes to use a chain of networks with increasing complexity, allowing simple inputs to complete processing with smaller networks than more complex ones. Evaluated on various classification tasks, they achieve 1.2-9.8x average reduction of operations per benchmark. The Envision NN accelerator [90] demonstrates this on a face recognition task, hierarchically increasing the network complexity, starting with a 12MOP network for presence detection (6.4mW, active 98% of time), followed by a network recognizing the owner, a set of 10 identities, 100 identities, and finally 5760 different identities (77mW, 0.01% of time).

#### 2) *Early exiting*

Adding special output classifiers after every few layers allows terminating a NN execution early if classifiers report a high confidence [131]. The trained network is analyzed after each layer to estimate a gain metric, quantifying the ratio between reduced number of operations and increased overhead due to the added classifier. Benchmarked on two 6- and 8-layer networks with 1 and 2 early exits, respectively, 1.73x and 1.91x reduction in number of operations are reported at iso accuracy. The same technique is used in the 12-class keyword spotting accelerator [132], reporting 69% of the inputs exiting early, reducing the average power consumption by 22% compared to always executing the complete network.

#### 3) *Selective execution*

Selective execution [133] enables different execution paths that can be selected depending on the input provided: an embedded selector network decides which branch to execute, providing less complex network branches for simpler inputs to reduce the average number of computations.

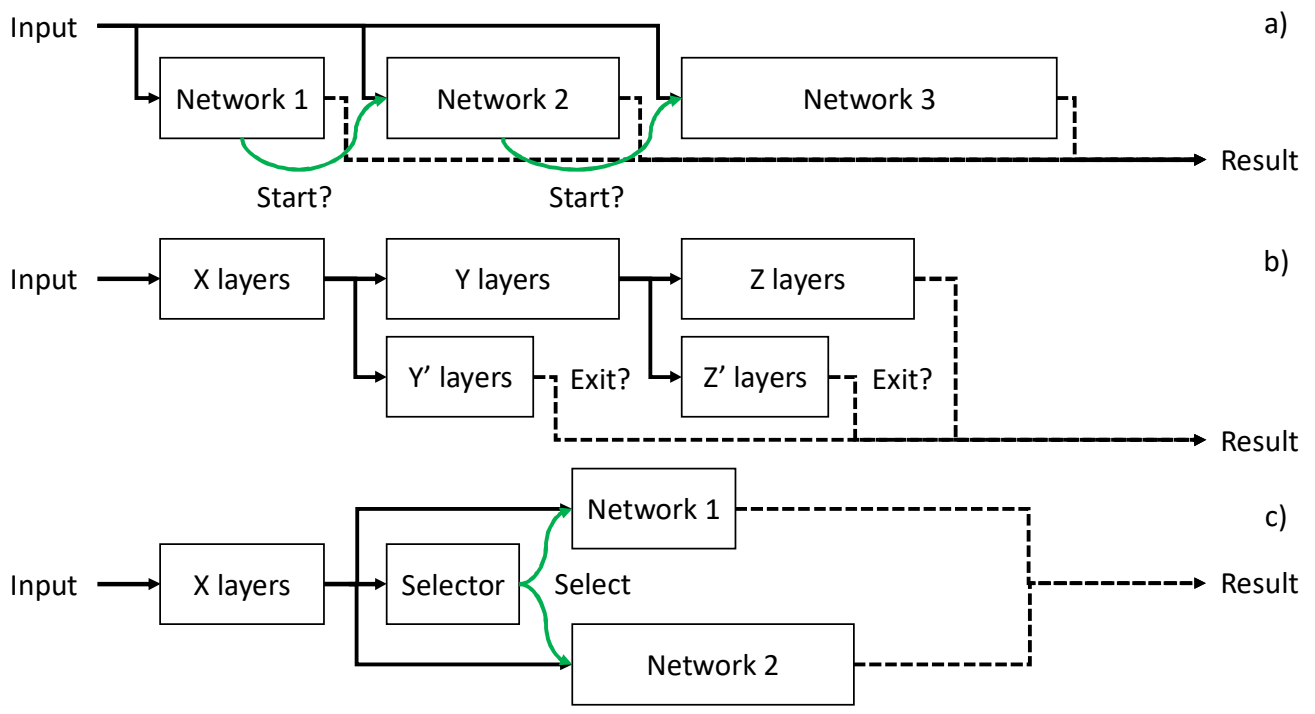

Fig. 7 Selective execution and early exiting approaches: a) scalable-effort execution, b) early exiting, and c) selective execution.

## VII. Data handling optimizations

The data-intensive nature of NNs challenges the memories and related access energy efficiencies. In many systems more than 50% of the total power consumption is related to memories and data handling [72]. This chapter discusses optimizations for efficient memory utilization and compression, reducing the amount of data to be accessed. Related sections cover the efficient reuse of data across computation elements (Section VIII.C) and the reduction of data through computational optimizations (Section VIII).

### A. *Efficient memory utilization*

Efficient memory utilization reduces the required memory space, saving power, chip area and thus IC cost.

The traditionally used buffering scheme for layer-wise NN processing is often called ping-pong buffering [120], following a double buffing approach to allow simultaneous reading of input activations and writing to output activations. It maps the activations of subsequent layers to two disjunctive memory regions, which must therefore allocate at least the maximum sum of any two subsequent layers. By allowing the activation memory regions to overlap during layer-wise processing, memory savings of up to 50% compared to standard ping-pong double buffering can be achieved [135]. The extent of savings depends on the layer dimensions and increases for large layers with small kernel sizes.

### B. *Data compression*

Compression reduces the memory footprint of data content and can be adopted in NN accelerator designs.

Run-length compression (RLC) of zero values is used in Eyeriss [136] to reduce the memory footprint and bandwidth. It encodes the number of zero entries in 5b, followed by the next non-zero value, reporting 1.2-1.9x reduced memory accesses for AlexNet. Eyeriss v2 [137] uses a "compressed sparse column format" for both weights and activations, allowing to skip sparse operations directly in the compressed form, reducing memory bandwidth and energy. Compared to RLC, it simplifies addressing of sliding window striding.

Loss-less Huffman coding [138], is shown to reduce weight memory by 20-30%. It employs variable-length codewords, providing smaller bit-widths for more common values, reducing the overall memory. An edge ML Huffman-coding DMA is shown to reduce data bandwidth by up to 5.8x [90].

Weight sharing is used in [138], replacing weights with table indices, referencing a limited number of pysically stored values. The upper bound of memory savings is defined by the weight bit-width $N_{width,w}$ and the number of table entries $N_{values}$ as shown in equation (2). The 45nm accelerator EIE [139] reports 8x energy savings through weight sharing (4b indices referring to 16 16b weight values).

$$N_{wshare,max} = N_{width,w} / \log_2(N_{values}) \tag{2}$$

## VIII. Computation optimizations

NNs contain millions of MAC operations, requiring fast and efficient accelerator designs. This chapter discusses computation optimizations ranging from operation reductions (optimized convolution operations, sparsity, or data reuse) to arithmetic simplifications (quantization, approximate computing, energy-quality scaling, or non-conventional arithmetic) and circuit optimizations (mixed-signal arithmetic, non-conventional arithmetic).

### A. *Optimized convolution implementations*

The dominance of convolution operations in many network architectures [140, 141] motivated optimized convolution implementations, aiming for similar algorithmic behavior while reducing computational complexity and resources.

#### 1) *Separable convolutions*

Separable convolutions are based on a separable filter approximation, splitting higher dimensional kernels (e.g. a $k_x \cdot k_y$ 2D convolution) into multiple lower-dimensional ones (e.g. 2 1D convolutions of $k_x \cdot 1$ and $1 \cdot k_y$), significantly reducing the number of operations.

A 2D approach is used in [106], replacing a 5x5 convolution layer with a horizontal and a vertical 5x1 and 1x5 layer, reducing the number of operations by 4x. Evaluated on an LFW face recognition NN, the total number of operations is reduced by 1.7x, while accuracy was decreased by 1%. To optimize parallel data access for the vertical direction, a transpose SRAM (T-SRAM) is proposed, enabling both row and column vector readout, reducing power by 47%.

Depth-wise separable convolutions (DSC) separate the kernel only in the depth dimension, convolving inputs in the spatial directions, followed by a pointwise (across depth) convolution that combines the filtered inputs to an output. In contrast to standard convolutions that combine the filtering and output generation, DSCs enable memory savings and reduces MAC operations. MobileNets [142] use such DSCs, reporting computation and parameter size reductions according to equation (3). Evaluated for the ImageNet task, parameters can be reduced by 7x while the number of MAC operations is reduced by 8.5x at an accuracy drop of just 1%.

$$reduction = \frac{1}{c_{out}} + \frac{1}{k_x \cdot k_y} \tag{3}$$

#### 2) *Frequency domain computation of CNN (FDC):*

Transforming a convolution operation into the Fourier domain results in a point-wise multiplication as shown in Equation (4) [143]. This property can be exploited to reduce the number of operations for computing CNNs. While the forward and backward transformations, using fast Fourier transformation (FFT), increase the computational effort and memory needs, the operation count can be significantly reduced for large input and kernel sizes, achieving 1.75-5.3x faster computation at iso-accuracy for 3x3 and 11x11 kernels in experiments on various layer sizes [143]. The reported speedup $s$, in terms of operation counts, is shown in Equation (5) for batch size $B$. However, this method requires large memories for the FFT and the subsequent matrix multiplication. A recent study [144] builds up on the FFT-based approach, additionally exploiting tiling, result-reuse, and symmetry of real-valued FFTs, roughly cutting the number of operations and Fourier outputs in half. It demonstrates 0.96-1.74x increase in throughput compared to the pure FFT-base approach on 9x9-3x3 kernel sizes.

$$f * g = \mathcal{F}^{-1}\{\mathcal{F}\{f\} \cdot \mathcal{F}\{g\}\} \quad (4)$$

$$s = \frac{(B \cdot c_{in} \cdot c_{out} \cdot x_{out}^2)}{(2 \cdot x_{out}^2 \cdot \log(x_{out}) \cdot (c_{out} \cdot B + c_{in} \cdot B + c_{in} \cdot c_{out}) + 4B \cdot c_{in} \cdot c_{out} \cdot x_{out}^2)} \quad (5)$$

#### 3) *Winograd algorithm*

For highly parallelized convolution computations, the sliding window operation can be flattened to convert it into a large point-wise matrix multiplication. The overlapping windows create significant redundancy with neighboring data, allowing to combine certain kernel-weights offline, which reduces the number of multiplications at the cost of more additions [141]. This so-called Winograd convolution achieves 1.48-2.26 speedup for computing 3x3 convolutions.

#### 4) *Strassen algorithm*

Large matrix multiplications, as used in NNs with large kernels, can be efficiently computed using the recursive Strassen algorithm, reducing the number of operations [145]. Evaluated on AlexNet, operations are reduced by 47%: while the dominating 3x3 and 5x5 convolutions are reduced in terms of operations, the 11x11 convolution in the first layer suffers from an increase of 18% compared to standard multiplication.

### B. *Sparsity exploitation and pruning*

Le Cun et al. [146] observed more than 3 decades ago that a significant number (75%) of NN parameters can be removed without affecting its algorithmic accuracy. A recent exhaustive survey [147] provides explanations to this phenomenon and estimates 10-100x model size reduction for various networks. It focuses on sparsification methods that set parts of a network to zero (pruning), while keeping its complexity constant. They reduce a network's size to its minimum required complexity by creating a (too) high dimensional representation to improve the training, knowing that the network can be reduced again through pruning. Previous works, report weight sparsity ratios of up to 99.996% for a LeNet-5 network with 99.3% accuracy on the MNIST task [148]. We refer the interested reader to the literature for more details on the main reduction techniques: model down-sizing through neural architecture search [149], operator factorization [150], quantization (Section VIII.D.3)), compression (Section VII.B), parameter sharing [151], and sparsification [147]. A survey on hardware acceleration of compressed models can be found in [152].

To create more sparse models, a 3-step approach is proposed [153], first learning the importance of each connection, then dropping low-weight ones and finally fine-tuning the training. Evaluated on AlexNet and VGG16 running the ImageNet task, they report 9x and 13x parameter reduction at iso-accuracy. Energy-aware pruning [154] optimizes the pruning strategy to achieve a minimum energy cost. Observing that layers which are pruned in an early stage, tend to have larger sparsity, they start pruning energy-intensive layers first, estimating their cost based on the number of computations and memory accesses. Evaluated on AlexNet running the ImageNet task, reports 3.7x energy reduction and 11x weight reduction at < 1% accuracy drop.

Envision [90] exploits sparsity by skipping sparse memory accesses and MAC operations using a sparsity map (1b entries per value). It reports 1.6x system energy saving for 30-60% activation sparsity. Similarly, NullHop [155] exploits sparsity by skipping sparse computations using a sparsity map combined with non-zero value list compression, achieving up to 3.68x throughput increase in 28nm synthesis results.

Zero-value skipping logic using a zero-free neuron array format is evaluated on a 65nm accelerator [156], reporting 1.37x average throughput increase for various networks at the cost of 25% memory increase (for zero sparsity).

A special form of temporal sparsity is proposed in CBinfer [157] and discussed in section VIII.C.

### C. *Data reuse*

Memory data that is used multiple times within a short period of time can be buffered in a local cache memory to allow cheaper and faster access. Special focus is set on data that has been fetched from energy expensive external memory. Three main structures of such data reuse have been studied extensively [27] and are summarized below, namely row-stationary (RS), weight stationary (WS) and output stationary (OS) approaches. While a WS setup minimizes movements of weights, as it is used in CIM architectures, OS keeps the partial sums for computing each output feature local, and RS approaches combine weight and activation data reuse.

Eyeriss [136, 50] implements a RS approach to maximize on-chip data-reuse and reduce costly external memory accesses. The 65nm chip can buffer one activations row (up to 224 16b values) and one weight row (up to 12 16b values), increasing energy efficiency by 1.4-2.5x.

A weight stationary approach is presented in [158], reducing the weight memory accesses by moving activations along cached weights during convolution computations.

OS processing is used in ShiDianNao [159], computing one output per processing element in its 8x8 array (64 outputs of the same output channel). This avoids moving partial sums to the memory and allows sharing inputs with neighboring PEs, reducing the memory bandwidth by up to 10x. Compared to their prior work DianNao [160], featuring no local data reuse, it allows to reduce power consumption by more than 1.66x, while increasing throughput by 1.87x. Another output stationary approach is used in Hyperdrive [161], keeping the feature maps stored entirely on-chip and streaming in weights.

This is motivated by the use of binary weights, consuming 16x less memory bandwidth than the 16b float activations.

Eyeriss v2 [137] implements a flexible network-on-chip (NoC) that can be reconfigured into 4 main reuse modes, enabling high activation and weight reuse in convolution layers while the dense layer mode can maximize activations broadcast because weights cannot be reused. Evaluated on MobileNet, they report a throughput increase of 5.6x.

Temporal data reuse is proposed in CBinfer [157], demonstrating that CNN-based CV applications with a static field of view can reuse large portions of each CNN layer and only compute those features that changed over time. They first detect changes in the input, generating a temporal sparsity map to update the connected output features for which the inputs exceed a calibrated threshold and buffer the new feature map. Evaluated on a 5-layer CNN for 10-class scene segmentation, this achieves an average speed-up of 8.6x.

### D. Hardware/Software co-optimization

A recent publication proposes less artificial intelligence [162], suggesting that today's networks are too high dimensional and thus prone to overfitting limited datasets, providing some intuition why high sparsity and quantization are viable optimization strategies. Today's NN algorithms exploit this observation, being developed with the challenges of resource-constrained edge ML hardware in mind. Thus, optimized algorithms like MobileNets [142] or SqueezeNets [163], reducing complexity through separable convolutions and kernel size minimization, have been introduced. These co-optimization strategies are covered in this section.

#### 1) Complexity scaling

The growing number of network architectures created a large design space, shifting the design strategies from hand-crafted architectures to (semi) automatic network architecture search (NAS), identifying optimal trade-offs between design constraints like accuracy and computational complexity. Frameworks like adaDeep [164] provide multi-dimensional model selection strategies to find the optimal model scaling strategies for a specific model and use-case. Dynamic complexity scaling is proposed by once-for-all networks [165], which are trained once but can then be deployed in various down-scaled complexities (in depth, width, kernel size, and resolution) without re-training. This allows deploying them on platforms ranging from high-performance cloud servers to low-power edge devices, with low accuracy degradations for the reduced-size versions.

#### 2) Energy-quality-scaling

Energy-quality (EQ) scalable systems [166] introduce a quality metric that describes a network's complexity. This allows to identify the knobs for power-scaling through (acceptable) quality degradations. The presented framework for EQ-scalable systems and EQ architectures [166] helps identifying applications where sensing and/or processing quality degradation is acceptable, for example in noise-resilient applications like computer vision. The list of quality knobs for dynamically adjusting EQ scaling encompasses arithmetic precision, bit error rates, sampling rates, algorithmic complexity and more. Follow-up works exploit this concept for implementing ULP voice activity detection [167] or always-on computer vision system [168]. Voice activity detection is shown to support EQ scaling [167], achieving 3.5x lower energy for 2% accuracy degradation using decision trees on a 28nm chip. Joint voltage and EQ scaling applied to a traditional computer vision task [168] is shown to achieve 3x lower energy through EQ scaling and 3.4x lower energy through VDD scaling on a 40nm process.

#### 3) Quantization and reduced precision

NNs can tolerate significant parameter and activation quantization with negligible effects on accuracy [40]. To reduce the effect of reduced precision, quantization-aware training techniques are used (e.g. straight-through estimators [169]) to keep the model derivable for back-propagation. Quantization works especially well in networks which are limited in training data, while the accuracy degradation increases for smaller (complexity-limited) networks, which cannot compensate for the loss of information [170]. Lowering precision reduces memory size, simplifies computational arithmetic, and lowers the power consumption, which also motivated Google to add 8bit support in their TPUs [49]. The viability of fully binary NNs (BNNs) [171], limiting activation and weight precision to 1bit, finally demonstrates the full range of quantization possibilities. BNNs have considerable accuracy degradations but allow to process multiplications using simple XNOR logic, reducing area and power needs.

Survey [41] provides an in-depth analysis of quantization schemes with a special section on sub-8bit quantization and a short overview of quantization-optimized hardware. Weight sharing is another form of quantization, limiting the number of supported values, as discussed in Section VII.B.

Energy and area savings of reduced precision arithmetic have motivated quantization optimizations. The energy for additions and multiplications are evaluated on a 45nm process [72], reporting 14x and 20x MAC energy reduction for 8b int compared to 32b int and 32b float, respectively, as summarized in TABLE III. A 45nm overview [172] reports power and area increase with bit-width for adders (linear increase) and multipliers (quadratic increase), as shown in TABLE IV. Comparing a MAC-combination, total area and power are reduced by 13x and 10.8x, respectively, for 32b to 8b, and by 3.6x and 3.6x, respectively, for 16b to 8b. Similarly, this was shown on multipliers for a 65nm process [160] as illustrated in TABLE V.

TABLE III
45NM ENERGY PER OPERATION FOR DIFFERENT PRECISION

| Precision | Int8 | Int32 | Float16 | Float32 |
|---|---|---|---|---|
| Addition energy [pJ] | 0.03 | 0.1 | 0.4 | 0.9 |
| Multiplication energy [pJ] | 0.2 | 3.1 | 1.1 | 3.7 |
| Total energy/MAC [pJ] | 0.23 | 3.2 | 1.5 | 4.6 |

TABLE IV
45NM ADDERS AND MULTIPLIERS FOR DIFFERENT PRECISION

| Precision | | Int8 | Int16 | Int32 |
|---|---|---|---|---|
| Adder | Area [$um^2$] | 212 | 322 | 1117 |
| | Power [uW] | 753 | 2235 | 4819 |
| Multiplier | Area [$um^2$] | 1038 | 4209 | 15126 |
| | Power [uW] | 2830 | 10816 | 34034 |

TABLE V
65NM MULTIPLIERS FOR DIFFERENT PRECISION

| Precision | | Int16 | Float32 |
|---|---|---|---|
| Multiplier | Area [$um^2$] | 1309 | 7998 |
| | Power [uW] | 577 | 4230 |

The 65nm accelerator in [160], achieves 6.1x reduction in area and 7.33x in power consumption for implementing 16b fixed point multipliers instead of baseline 32b floating point arithmetic while maintaining comparable accuracy. Envision [90] exploits 1-16b dynamic precision scaling combined with DVFS, reporting reductions in energy per MAC operation (relative to 16b) of >5x for 8b and >50x for 4b precision using sub-word parallel computations in a 28nm process.

UNPU [173] employs a non-linear quantization support that replaces a 16b multiplier by a 2x4-bit lookup table, indexing 16 16b activations and 16 16b weights, with the result of each combination stored in the table. They report 79% power reduction and 93% lower latency while the area is reduced by 1.3x compared to instantiating a 16b multiplier.

A review on scalable precision MAC architectures [174] compares recent 2-8b scalable implementations, discussing spatial and temporal MAC architectures and benchmarking them on a 28nm process using a data-gated 8b-input MAC as baseline. Throughput is roughly increased quadratically for cutting precision by a factor of 2, reaching up to 14.5x for 2b precision. Area is increased 1.1-4.4x for parallel precision-scalable designs while bit-serial implementations can reduce area by up to 40%. The overhead for scalability-support increases the energy per operation in full precision modes by up to 52% for single level, up to 94% for dual level scalability, and up to 14x for multi-cycle bit-serial MACs. The energy per operation, only reduces linearly with precision in the baseline but decreases supra-linearly for the scalable MACs, achieving up to 4x lower energy at 2b precision (overhead compensated at 6-4b precision for parallel and at 2b for bit-serial MACs).

An XNOR-based 22nm BNN accelerator [108] exploits the reduced precision by utilizing SCM instead of SRAM, enabling lower power consumption. Similarly, the 65nm binary-weight (12b activation) CNN accelerator YodaNN [175], reports improved performance using voltage-scalable SCM. Combining binary weights (instead of 12b) with SCM (replacing SRAM) allows them to reduce the power by 11.6x.

Cross-layer bit-width optimization [176], shows more than 20% parameter size reduction compared to homogeneous bit-width fixed-point quantization at iso-accuracy on the CIFAR-10 task. Furthermore, knowledge distillation can be used for low-precision quantization [177], improving the accuracy of a highly quantized model using "distilled" knowledge from a larger (higher precision) teacher network during training.

### E. Approximate computing

Approximate computing trades power consumption, speed, and area off against arithmetic accuracy [178]. Approximation approaches are either based on voltage over-scaling (VOS) below the technology's threshold voltage or on functional modifications ranging from algorithm- to circuit-level [179]. It differs from energy-quality-scaling (Section VIII.D.2), due to its circuit-based scaling approach.

A 2020 survey [180] on approximate computing for DNNs reports power, delay and area numbers from synthesized approximate adders, multipliers and dividers using a 28nm process at 1V. It reports up to 69% energy savings (power-delay product) for an image sharpening task while for JPEC compression only 20% savings are achieved.

An earlier survey [179] reports an approximate integer data format and related arithmetic operation implementations in 45nm [172], limiting values and computation precisions to a dynamically selected range of most significant non-zero bits. This achieves 55-65% power reduction compared to accurate computations at <0.5% accuracy drop in KNN and SVM tasks. Approximate computing using 2- and 3-bit adder designs [181] reports further power and accuracy improvements.

VOS introduces bit errors due to missed timing constraints and other unwanted effects but reduces power consumption as shown on a 28nm CNN accelerator [182]. Reducing the SRAM voltage from nominal 1.0V to 0.51V enables 3.12x memory and 2.13x system power reduction for running a 9-layer fully binary CNN at <1% accuracy drop. They report stronger effects on accuracy from weight errors than activation errors, enabling further activation memory voltage scaling.

### F. Non-conventional arithmetic

To further optimize NN computations, non-conventional computer arithmetic has been surveyed [183], comparing currently used CMOS technology and alternative emerging technologies for implementing computer arithmetic. Also alternative number systems are evaluated, for example a logarithmic system on a 65nm process [184], showing 3x higher energy per addition compared to floating point, but 1.5x lower for multiplications, 17x lower for divisions and 38x lower for square root computations.

#### 1) Spiking arithmetic

Biological neurons in the human brain function with spike-based signaling and computing, inspiring researchers to rethink the traditional level-based arithmetic in ICs [185]. Neuromorphic spiking arithmetic is employed in IBM's 28nm TrueNorth [186], implementing a total of 1 million digital spiking neurons, and Intel Loihi [187], implementing 131k neurons in a 14nm process. Due to the significantly different computing paradigm, which cannot be directly compared with other optimization approaches, we refer the reader to the specific literature for more details [188, 189, 185].

#### 2) Hyperdimensional computing

Hyperdimensional computing is another brain-inspired computing approach that encodes information in very high-dimensional binary vectors with thousands of entries, called hypervectors. The similarity of information contained in hypervectors is encoded in a distance metric, making them robust against bit errors and thus suitable for VOS. Tasks like image recognition are performed by comparing the distance of an input vector (e.g. features of an image) with a known reference vector. We refer to the specific literature [183, 190] for more details as this goes beyond scope of this work.

### 3) Quantum computing

While still being in its infancy, quantum computing [191] promises extremely powerful computing capabilities, enabling unprecedented throughputs which could allow further acceleration of NN computations in the future.

## G. Mixed signal arithmetic

The analog nature of most sensor signals and power-advantages of computing in the analog domain motivated mixed signal arithmetic for computing DNN [192, 193].

Survey [34] compares a set of analog DNN accelerator architectures, reporting 40-80% lower area as well as 70% and 40% reduced power compared to digital implementations for 130nm and 65nm designs, respectively. This shows that analog circuits do not scale equally well with reduced process nodes as their digital counterparts. Other properties, like the intrinsic computation parallelism from Kirchhoff's law can compensate for this reduce advantage in smaller node sizes.

Analog implementations usually require peripheral circuits that can diminish analog computation advantages at increasing design efforts, as illustrated in [194, 195], implementing the same accelerator in a 28nm process but using analog or digital MAC-accumulation circuits, respectively. Evaluated on a fully binary CIFAR-10 network at iso-accuracy, the energy per inference dropped from 14.4uJ (digital) down to 3.79uJ (analog), which is a system-level energy improvement of 3.8x, while the energy for the underlying MAC computation dropped by nearly 12.9x (>3x more). The 28nm analog-domain computations (8b dot product) are presented in [196], reporting nearly 75% energy spent on ADC and control logic.

Bong et al. [197] implement a hierarchical analog-digital hybrid binary decision tree engine on a 65nm image sensor, running 60% of the algorithm in the analog domain, reducing the inference energy by 39% compared to digital computation.

A 130nm 32x32 analog MAC array multiplying a DAC-converted vector with a 32x32 matrix is presented in [198]. Compared to a multi-core processor baseline, power and area are reduced by 71% and 43%, increasing throughput by 10.3x.

The CIM approaches discussed in Section IV.C exploit the advantages of mixed signal processing, keeping the data in memory to avoid losses data movement and digitalization losses. An RRAM-based analog crossbar [199], compares performance to equivalent implementations with digital RRAM-usage and SRAM-based memory. It reports 270x energy and 540x latency improvements over digital RRAM, and 430x energy and 34x latency improvements over SRAM implementations, showing latency issues for digital RRAM.

## H. Arithmetic implementations

NN training is usually executed with 32bit floating point precision, that can be significantly lowered during inference. Each layer has a specific sensitivity to quantization and can thus be implemented with adapted (minimal) bit-widths [200]. Selecting the arithmetic precision and the data types allows to optimize implementations, increasing throughput and energy efficiency as shown in Section VIII.D.3).

Motivated by the finding that the required precision varies across DNN layers [200], a 65nm bit-serial DNN engine is implemented based on the 16bit DaDianNao architecture [201]. Evaluated on 9 common DNNs with per-layer-minimized precision, it reports 1.2x-4.76x (2.0x on average) increased energy efficiency at iso-accuracy compared to the baseline accelerator.

Variable-precision bit-serial MAC can also be implemented using look-up tables [202], reporting energy savings of 23%, 27%, 41% and 54% with respect to standard fixed-point MAC for 16-, 8-, 4-, and 1-bit weight precision, respectively (16b activation). A similar, Booth-Wallace multiplier-based multi-precision implementation was shown in [90], supporting 16b multiplications, that can be split into 4 x 4b multiplications.

# IX. Quantitative Comparison of Optimizations

This chapter summarizes the quantitative effects of the discussed edge ML accelerator optimizations. TABLE VI - TABLE VII list each optimization approach with a brief description of the implementation setup that was used to demonstrate the effect in the referenced publication. Five performance indicators quantify each technique: the memory usage impacts the (often dominating) energy for data handling, the throughput determines the processing latency, the chip area directly impacts manufacturing cost, and the power/energy reductions translate into longer battery lifetimes. However, optimizations might influence the algorithmic accuracy, which is therefore listed in the fifth impact column.

Architectural optimizations (Section IV) report up to 13x power savings with increased throughput using CIM. Note that currently, most implementations only support small networks. Power management (Section V.B) allows for significant leakage and dynamic power reductions, but negatively impacts the throughput due to the reduced operating frequency. Optimizing the memory offers reduced access energy and mainly lower static power, while affecting the required area. Optimized placement and low-leakage SRAM types allow trading area off against power. Various dataflow and data handling options (Sections VI-VII) offer improved throughput or reduced memory requirements (up to 1'000'000x lower size). Computation improvements (Section VIII) higher throughput and efficiency using quantization but must be traded off against accuracy deteriorations.

As an example, the designer of a new edge ML accelerator, requiring to run low-complexity ML tasks at a very low power consumption, could identify DVFAS [90] in TABLE VI as useful optimization, reporting 25x power reduction at constant throughput. The reported accuracy impact must be considered but might be acceptable for simple tasks. To further improve the power consumption, SCM can be chosen instead of standard SRAM, adding another 2-3x reduction in (memory) power consumption [108]. If the selected network tolerates sparsity, the support of weight and activation sparsity from TABLE VII might be another option to drastically decrease the power further. However, replacing the digital processing with a mixed-signal implementation from TABLE VII promises lower power gains and probably comes at the cost of increased area, as reported in the literature [194].

TABLE VI
OVERVIEW OF OPTIMIZATION STRATEGIES AND REPORTED ADVANTAGES

| **Field (chap.)** | **Optimization approach** | **Reported impact** | | | | | **Reference work** | | **Remarks** |
|---|---|---|---|---|---|---|---|---|---|
| | | **Memory usage red.** | **Throughput increase** | **Area reduction** | **Power/Energy reduction** | **Algorithmic** | **Work (Year)** | **Implementation performance** | |
| **Architecture (IV)** | Replace systolic array PEs with tensor PEs | - | - | 2.1x | 1.4x system | - | [51] (2020) | 16nm accelerator running various nets (e.g. ResNet-50, MobileNetV1) | |
| | Replace MAC with CIM-MAC | - | increased | unknown | 13x system | impacted (quantization) | [57] (2017) | 128x128 CIM array in 130nm running MNIST task (1b w. , 5b act.) | Tiny network presented only |
| | Replace MAC with SRAM CIM-MAC | - | 2.4x | unknown | 4.9x system | negligible | [58] (2018) | LeNet5 network running MNIST task on simulated 65nm accelerator | |
| | Add 8b CIM to pure SRAM | - | increased (none before) | 0.78x (28% larger) | unknown | - | [59] (2021) | 384kB SRAM with 8b CIM in 28nm process | |
| **Power management (V.B)** | Sub-threshold operation | - | 0.23x (4.4x slower) | 0.53x (1.87x larger) | 5x power | - | [82] (2013) | 180nm sub-threshold standard cell compared to standard 180nm library | |
| | Sub-threshold operation | - | reduced | - | 7x power | - | [84] (2016) | 180nm MCU: 13kHz @ 0.48V vs. 25MHz @ 1.8V | |
| | Sub-threshold operation | - | reduced | - | 3.8x power | - | [83] (2015) | 180nm 1kb SRAM @ March test: 530Hz @ 0.27V vs. 200kHz @ 0.6V | |
| | ABB | - | Reduced (30x lower) | - | ≤ 20x leakage | - | [85] (2019) | 55nm RISC MCU and SRAM | |
| | DVFS | - | - | - | 1.05-4x | - | [87] (1994) | CPU @ varying intensity using DVFS | |
| | DVFS | - | - | - | 9.2x | - | [88] (1996) | CPU @ 1.5V and 1/10 frequency DVFS vs. CPU @ 3V and 90% idle | |
| | DVFAS | - | - (constant) | - | 25x | Impacted (quantization) | [90] (2017) | 28nm acc.: $4 \cdot 4b$ with DVFAS vs. $1 \cdot 16b$ @ constant throughput | DVFAS: 0.65-1.1V, (BB: 0.2-1.2V) |
| | DVAS | - | Beneficial | 0.85-0.9x (<15% more) | 11.1x(16b -> 8b) 0.8-0.9x (@16b) | Detrimental | [89] (2015) | 40nm 16b Baugh-Wooley multiplier with DVAS vs. gating unused bits | Energ. overhead due to add. logic |
| **Memory (V.C)** | Reduce memory size for lower access energy | impacted | - | with memory | Linear with memory size | - | [91] (2016) | 45nm SRAM implementations: 1-1024kB | Non-linear for small sizes |
| | Replace 6T SRAM with FeFET | - | 1.5x reading 0.1x writing | >5.3x | 74x static power, 0.6/0.1x read/wr. | - | [103] (2019) | 22nm 64x64b array: FeFET vs. 6T SRAM (FeFET read/write: @4V/1V) | Read/write energy increases |
| | Reduce SRAM area: low leak. 7T vs. 8/10T | - | similar | 18% 50% | similar | - | [17] (2019) | 8kB SRAM in 180nm using 7T vs. 8T and 10T low leakage SRAM | |
| | Non-uniform SRAM instead of uniform | - | - | 0.98 (2% larger) | 60% (1.5kB vs. 48kB access) | - | [104] (2017) | 4-level SRAM memory (1.5-48kB) in 40nm at 0.65V and 3.9MHz | |
| | ULP SRAM sleep mode | - | Detrimental | 0.37x (2.7x larger) | 26x | - | [105] (2020) | 16kB SRAM with charge pump for low power retention in 55nm | Slow process corner is used |
| | Reduce SRAM leakage 10T vs. 8T | - | unknown | 0.09x (11.3x lager) | 4242x retention | - | [73] (2013) | Low leakage 180nm 10T SRAM vs. 65nm 8T SRAM | |
| | Transpose SRAM vs. row-only SRAM | - | - | Detrimental (added logic) | 1.9x | . | [106] (2018) | 65nm accelerator with T-SRAM vs. SRAM for 5x1 & 1x5 sep. conv. | |
| | Replace SRAM with SCM | - | similar | 2-3x (<1kb), <1/3x (>1kb) | >2x | - | [107] (2016) | 8kb SCM vs. SRAM in 28nm using dedicated SCM placement | |
| | Optimized placement strategy for SCM | - | similar | >35% | Read: ≤65%, Write: ≤50% | - | [107] (2016) | 256b-32kb SCM macros in 28nm: special placement vs. standard | |
| | Replace SRAM with SCM | - | - | Not reported | 2-3x | - | [108] (2018) | BNN accelerator in 22nm using SRAM@0.6V or SCM@0.4V | |
| | 3D HMC DRAM instead of 2D DRAM | - | 4.1x | Beneficial (3D-stacked) | 1.5x | - | [110] (2017) | HMC DRAM dies on 45nm logic die running various networks | |
| | Replace SRAM with eDRAM (4T) | - | - | 17% | 23% | - | [111] (2018) | 28nm 8kb eDRAM array vs. SRAM (both@0.7V and room temperature) | |
| | Replace ULP SRAM with eDRAM (2T) | - | unknown | 2.5-3.8x | Same leakage | - | [112] (2019) | 28nm 4kb eDRAM memory array at 0.4Vvs. SRAM | Simulation results only |
| **Dataflow & Control (VI)** | Depth-first instead of layer-wise processing | Size:≤ $10^6$x BW:≤200x | | Linked to memory | - | - | [115] (2019) | Mem. reduct.: e.g. SRGAN (UHD): 19'633x[1], MobileNetV2: >20x[1] | Limits number of supported layers |
| | Depth-first instead of layer-wise processing | on-chip:0.8x Ext.: 20x | (0.94x) | - | Beneficial (less ext. memory) | - | [117] 2016) | VGGNet-E on FPGA | AlexNet: only 28% ext. reduc. |
| | Optical convolution implementing 1st layer | 2.5x – 10x (E)/MNIST | 250x-460x (E)/MNIST [2] | - | Linked to throughput | -1.7% /-0.4% (E)/MNIST | [125] (2020) | CNN layer in front of 4 dense layers on MNIST and EMNIST task | Only works on first layer. |
| | In-sensor processing of first CNN layers | - | - | - | 73/85/45% sys./ sensor/compute | Detrimental | [127] (2016) | Sim. 180nm SAR ADCs and tunable cap. running ConvNet | first 1-5layers of 8b GoogleNet |
| | Add hierarchically scalable effort | Detrimental (more nets) | Beneficial | - | ~10x | - | [90] (2017) | 28nm DNN accelerator running 12MOP – 30.8GOP networks | |
| | Add hierarchically scalable effort | Detrimental (more nets) | 1.2-9.8x (avg.) | - | Linked to throughput | - | [134] (2015) | Hier. scaled SVM, NN, and dec. tree on MNIST and other tasks | Overhead for difficult inputs |
| | Introduce early exit for conditional execution | Detrimental (new layers) | 1.73-1.91x | - | Linked to throughput | Slightly increased | [131] (2016) | 6- and 8-layer NN for MNIST task with 1 and 2 early exits | |
| | Introduce early exit for conditional execution | Detrimental (new layers) | - | - | 1.22x | Slightly reduced | [132] (2020) | 12-class Google speech command task with early exit on 22nm acc. | |
| **Data handling (VII)** | Replace ping-pong buffering | Up to 2x | - | - (with memory) | - | - | [135] (2020) | Act. (total) memory reduction e.g. DMCNN-VD: 48.8% (48.2%) | |
| | Compression using Huffman coding | 1.2-1-3x for weights | - | - (with memory) | - | - | [138] (2016) | Encode weights using Huffman coding in AlexNet and VGG16 | |
| | Compression using Huffman coding | Up to 5.8x | - | - (with memory) | - | - | [90] (2017) | 28nm DNN accelerator running face recognition CNNs | |
| | Compression using RLC coding | 1.2-1.9x | - | - (with memory) | - | - | [136] (2016) | 65nm process running AlexNet | |
| | Weight sharing | <4x for weights | - | - | 8x | Similar | [139] (2016) | 45nm accelerator with weight sharing (16 entries of 16b values) | |

[1] at equal memory bandwidth (lower for lower memory bandwidths) [2]with respect to largest network with similar performance

TABLE VII
OVERVIEW OF OPTIMIZATION STRATEGIES AND REPORTED ADVANTAGES

| Field (chap.) | Optimization approach | Reported impact: Memory usage red. | Throughput increase | Area reduction | Power/Energy reduction | Algorithmic | Reference work: Work (Year) | Implementation performance | Remarks |
|---|---|---|---|---|---|---|---|---|---|
| Computation (VIII.A) | Replace CNN with depth-wise separable conv. (MobileNet) | >7x | 8.5x | Linked to memory | - | 1% | [142] (2017) | MobileNet vs. pure CNN on ImageNet task | |
| | Separable convolution instead of 5x5 (1x5, 5x1) | Beneficial (less param.) | 4x (1.7x on total network) | - | - | -1% | [106] (2018) | 5-layer CNN for LFW face recog. with sep. conv. vs. normal conv. | |
| | FFT-based convolution | Detrimental | 1.75-5.3x | - (with memory) | Linked to throughput | - (iso-accuracy) | [143] (2014) | Computing 3x3 – 11x11 CNN kernels via FFT vs. conventional | Smaller kernels profit less |
| | Optimize FFT-based conv. using fine-grained FFT | Beneficial | 0.93-1.74x | - | Linked to throughput | - | [144] (2020) | Fine-grained FFT-based conv. vs. pure FFT conv. (9x9–3x3 kernels) | |
| | Use Winograd fast convolution | - | 1.48-2.26x | - | - | Negligible normally | [141] (2016) | Winograd 3x3 convolution on VGG E network (batch size 64-1) | Accuracy drop for large kernels |
| | Use Strassen algorithm for matrix multiplication in CNNs | - | 24-47% | - | - | - | [145] (2014) | Strassen algo. on AlexNet conv. layers 2-5 (5x5, 3x3) on CPU | More effective for lager matrices |
| Computation (VIII.B) | Sparsity exploitation + Pruning approach | 9x-13x for weights | Beneficial | - | - | - (iso-accuracy) | [153] (2015) | Network pruning on AlexNet and VGG16 for ImageNet | |
| | Sparsity exploitation: skip memory access and MAC op. | - | Beneficial | - | 1.6x | - | [90] (2017) | 28nm DNN accelerator with 1b sparsitiy map (16b activations) | Assumes 30-60% sparsity |
| | Sparsity exploitation with energy-aware pruning | 11x weights | unknown | - (with memory) | 3.7x | <1% | [154] (2017) | AlexNet on ImageNet task | |
| | Skipping sparse operations | Beneficial (sparsity) | 3.68x | - | - | Impacted (quantization) | [155] (2019) | 28nm CNN acc. with sparsity map and NZVL compression | Simulation only |
| | Add zero-value skipping logic | Beneficial (1.25x more) | 1.37x | Similar (adds logic) | Beneficial | - | [156] (2016) | 65nm acc. with zero-skipping logic and zero-free data encoding | |
| Computation (VIII.C) | Add data reuse: row-stationary processing | Detrimental | Beneficial | Detrimental | 1-4-2.5x | - | [136] (2016) | 65nm 168 PE row-stationary accelerator running AlexNet | Data in SRAM and ext. DRAM |
| | Add data reuse: output stationary | Detrimental (local mem.) | 1.87x | 0.3 (5.5x larger) | 1.66x | - | [159], (2015) | 65nm 64PE OS acc, with vs. without reuse on various CNNs | Savings depend on network |
| | Add data reuse: NoC for flexible reuse modes | Detrimental | 5.6x | Detrimental | 1.8x | - | [137] (2019) | 65nm NN accelerator with 192 PEs running MobileNet | |
| | Data reuse: change-based temporal sparsity | Detrimental | 8.6x | - (with memory) | - | - | [157] (2017) | On 5-layer CNN for 10-class scene segmentation | Entire net stored in memory |
| Computation (VIII.D) | Reduce precision int32/float32 to int8 | Beneficial | Unknown | Unknown | 14x (int32) 20x (float32) | Impacted | [72] (2014) | 45nm arithmetic compared | |
| | Reduce precision int32/int16 to int8 | Beneficial | Unknown | 13x (int32) 3.6x (int16) | 10.8x (int32) 3.6x (int16) | Impacted | [172] (2017) | 45nm arithmetic compared | |
| | Reduced precision (16b -> 8/4b) | Up to 4x | Up to 4x | - | >5x (16b -> 8b) >50x (16b->4b) | Accuracy impacted | [90] (2017) | 28nm acc. with sub-word parallel 1-16b MAC precision | |
| | Replace 16b multiplier with 4bx4b lookup table | - | 14.3x | 1.3x | 4.7x | Impacted (quantization) | [173] (2018) | 65nm acc running ImageNet task (quantized vs. 32b baseline) | Table: 256 pre-comp. 16b results |
| | Replace MAC with scalable precision MAC (8b -> 8/4/2b) | - | Up to 14.5x (2b) | >0.22x par.) < 1.4x (ser.) | up to 4x (2b) | Accuracy impacted | [174] (2019) | 28nm scalable prec. MAC designs (parallel and serial) vs. 8b MAC | 2x-14x power overhead @ 8b |
| | Reduce weight precision 12b -> 1b, replace SRAM->SCM | Weights: 12x | similar | 1.2x | 11.6x | Impacted | [175] (2018) | Binary weight accelerator in 65nm with SCM | |
| | Energy-quality scaling | - | - | - | 3.5x | 2% reduction | [167] (2020) | Voice activity detection using decision trees on 28nm chip | |
| | Energy-quality scaling: using det. thresh., feature/mem. size | - | - | - | 3x | Negligible | [168] (2020) | 40nm image feature detection | |
| Comput. (VIII.E) | Approx. computing | - | - | Beneficial | 20%-69% (pow. delay product) | Accuracy impacted | [180] (2020) | 28nm approx. mult./adders run. image sharpening, JPEG compr. | |
| | Approx. computing | - | - | Beneficial | 55% (KNN)- 65% (SVM) | Accuracy impacted | [172] (2017) | 45nm approx. comp. engine vs. accurate 32bit implementation | |
| | Approx. computing: SRAM VOS | - | 0.095x (10.5x slower) | - | 3.12x | Impacted (<1% lower) | [182] (2018) | 28nm fully binary CNN acc. running 9-layer CNN | |
| Computation (VIII.G) | Replace digital with mixed-signal arithmetic | - | 1.58x | 0.31 (3.2x larger) | 3.8x (system) 12.9x (MAC) | - (iso-accuracy) | [194], (2018) | 28nm fully binary CNN acc. with analog MAC vs. digital [195] | Analog periph. reduce gain >3x |
| | Replace float arithmetic with logarithmic | - | unknown | unknown | 0.33x,1.5x,17x (add,mult,div.) | | [184] (2016) | 65nm logarithmic arithmetic system vs. conventional float | |
| | Replace digital with mixed-signal arithmetic | - | - | - | 1.39x | Impacted | [197] (2017) | 65nm mixed analog/digital comp. vs. pure digital face detection | 60% of algo. in analog domain |
| | Replace digital with mixed-signal arithmetic | - | 10.3x | 1.7x | 3.4x power | - | [198] (2011) | 130nm 32x32 analog MAC array for matrix-vector multiplication | Input/output in digital domain |
| | Replace dig. SRAM MAC with analog RRAM- CIM | - | 34x | 11x | 430x | Deteriorate significantly | [199] (2018) | Simulated accelerator design in 16nm, running MNIST task | |
| Comput. (VIII.H) | Replace digital with mixed-signal arithmetic | Beneficial | 1.3-5.3 | 0.95x | 1.2x-4.76x (2.0x average) | - (at iso-accuracy) | [200] (2017) | 65nm var. perc. bit-serial acc. vs. 16b baseline | Evaluated on 9 common DNNs |
| | Replace MAC with variable prec. LUT-based bit-serial MAC | - | - | - | 1.23x-1.54x (16b-1b weight) | - | [202] (2019) | 65nm NN acc. with var. precision MAC vs. fixed 16b MAC | Activations: 16b |

## X. Conclusion

We presented a survey of design optimization strategies for low power neural network accelerators. The compiled list of optimization approaches provides quantitative performance impact measures for each approach, allowing accelerator designer to estimate their impact in future designs.